%% file: main.tex
\newif\ifanon \anonfalse
\newif\iffull \fullfalse

\documentclass[fleqn]{article}
\usepackage{ijcai18}

\usepackage{times}
\usepackage{xcolor}
\usepackage{soul}
\usepackage[utf8]{inputenc}
\usepackage[small]{caption}
\usepackage{microtype}

\usepackage{latexsym}

\input{rule}

\title{Profile-guided memory optimization for deep neural networks}

\ifanon
\else
\author{
Taro Sekiyama\thanks{This work was partially done in IBM Research - Tokyo.} \\National Institute of Informatics\\sekiyama@nii.ac.jp
\And
Takashi Imamichi\\IBM Research -- Tokyo\\imamichi@jp.ibm.com
\AND
Haruki Imai\\IBM Research -- Tokyo\\imaihal@jp.ibm.com
\And
Rudy Raymond\\IBM Research -- Tokyo\\rudyhar@jp.ibm.com
}
\fi

\begin{document}

\maketitle

\input{sec/abstract}
\input{sec/intro}
\input{sec/relwork}
\input{sec/approach}
\input{sec/impl}
\input{sec/exp}
\input{sec/conclusion}

\bibliographystyle{named}
\bibliography{main}

\end{document}

%% file: rule.tex
\usepackage{algorithm, algpseudocode}
\usepackage{amsmath}
\usepackage{amssymb}
\usepackage{graphicx}
\usepackage{subcaption}
\usepackage{url}

\newcommand{\sect}[1]{Section~\ref{sec:#1}}
\newcommand{\fig}[1]{Figure~\ref{fig:#1}}

\newcommand{\id}{\ensuremath{\lambda}}
\newcommand{\TODO}[1]{\textcolor{red}{TODO: #1}}
\newcommand{\TS}[1]{\textcolor{red}{TS: #1}}

\newcommand{\cpp}{C\nolinebreak\hspace{-.05em}\raisebox{.4ex}{\bf +}\nolinebreak\hspace{-.10em}\raisebox{.4ex}{\bf +}}

%% file: sec/abstract.tex
\begin{abstract}
 Recent years have seen deep neural networks (DNNs) becoming wider and deeper to
 achieve better performance in many applications of AI.
 Such DNNs however require huge amounts of memory to store weights and
 intermediate results  (e.g., activations, feature maps, etc.) in propagation.
 This requirement makes it difficult to run the DNNs on devices
 with limited, hard-to-extend memory, degrades the running time performance, and
 restricts the design of network models.
 %
 %
 %
 We address this challenge by developing a novel profile-guided memory
 optimization to efficiently and quickly allocate memory blocks during
 the propagation in DNNs.
 The optimization utilizes a simple and fast heuristic algorithm based on the
 two-dimensional rectangle packing problem.
 Experimenting with well-known neural network models, we confirm that our method
 not only reduces the memory consumption by up to $49.5\%$ but also accelerates
 training and inference by up to a factor of four thanks to the rapidity of the
 memory allocation and the ability to use larger mini-batch sizes.
\end{abstract}

%% file: sec/intro.tex
\section{Introduction}

Since its great success in computer vision~\cite{alexnet}, deep learning, the
machine learning technology based on \emph{deep neural networks} (DNNs), has emerged widely
in image processing, machine translation, speech recognition, and many AI applications.
One factor leading to its popularity is the increase of computational power.  Especially,
the effective use of graphical processing units (GPUs) has made it
possible to train sophisticated DNNs on huge datasets~\cite{alexnet,DL-Nature}. Along with the progress of research
to obtain better accuracy, DNNs
are becoming {\textit{deeper}}, i.e., having more layers, and/or {\textit{wider}}, i.e.,
having more branches and more neuron units in each layer. For example, in image
recognition, AlexNet~\cite{alexnet}, the winner of ILSVRC 2012, consists of only nine
layers and has a sequential structure; GoogLeNet~\cite{googlenet} introduces the
so-called inception modules, which are a technique to widen DNNs;
ResNet~\cite{resnet} consists of more than 50 layers; and
the more recent network, Inception-ResNet~\cite{incept-resnet}, which extends ResNet with the inception
modules, is even larger than ResNet and GoogLeNet.



Although expanding neural networks seems to be a key to obtain better accuracy, it
comes with the high memory cost required to store weight parameters and
intermediate results (e.g., activations, feature maps, etc.) in the propagation for
training and inference. For example, the training of Inception-ResNet
consumes 12.5 times as much memory as that of AlexNet in some configuration
(see \fig{exp-mem}). This gives rise to several undesirable consequences.
First, for training DNNs, we can only use smaller mini-batches to avoid risk running out of memory, and
hence having slow convergence. Second, inference using such large DNNs may require
many machines in the deployment environment.
%
%
Third, and even worse, the flexibility
of the design of neural networks is constrained so that the DNNs can fit in the
memory of the underlying devices. The high memory consumption is more serious in the use of GPUs
and edge devices that have much smaller and less extendable memory storage than CPUs{}.\footnote{
Augmenting devices may mitigate the problem but it makes another issue about the communication between devices.}
%

We study memory optimization for DNNs.
Our approach is based on the observation that propagation of a network model is computed in the
same way for different inputs and learnable parameters; we call such
propagation \emph{hot}.\footnote{This term originates from just-in-time
compilation, where repeatedly executed code blocks to be optimized are called
hot.}
This is indeed the case in
many networks including convolutional neural networks (CNNs) such as AlexNet, GoogLeNet, ResNet,
and Inception-ResNet. On the basis of this observation, we can profile the memory usage in a sample run
and then utilize the profile to find the allocation of memory to minimize
the peak memory usage in the succeeding runs. The allocation problem is a special
case of a two-dimensional rectangle packing problem that is known to be NP-hard.
We develop a simple and fast heuristic to generate good approximate solutions quickly.

Even when the \emph{whole} propagation is not hot,
there are cases where some \emph{part} of the propagation is hot. For example,
a recurrent neural network (RNN) with long short-term memory (LSTM) units~\cite{lstm}
takes variable-length inputs, and a part of its propagation is computed differently depending on inputs.
We find that our approach can also reduce the memory consumption in such RNNs, as discussed in
\sect{impl-gen}.

A byproduct of our approach is that training and inference of DNNs
may be accelerated.  After a sample run, our method produces an absolute memory
address for each memory request and just returns it in the succeeding runs.  The
run-time cost of our memory allocation is thus very low in the succeeding runs,
compared with that of dynamic memory allocation, a standard approach in many
deep learning frameworks.  We describe the dynamic method in \sect{relwork}
briefly and compare it with our approach from the perspective of the running
time performance as well as the memory consumption in \sect{exp}.
In addition, the memory reduction allows use of larger mini-batch sizes, which
leads to higher utilization of GPU cores and further acceleration of training.

Our contributions are summarized as follows.
\begin{itemize}
 \item We propose a profile-guided memory optimization technique for DNNs.
       Our approach optimizes memory usage in a hot part of a propagation and
       never incurs performance overhead once the memory usage is optimized,
       while preserving the computation of the DNNs.

 \item We develop a simple heuristic algorithm for allocating memory to minimize the peak memory usage.
       We empirically show that it works well from the perspectives
       of both computation time and solution quality.

 \item We implement our memory optimization on the common deep learning
       framework Chainer~\cite{chainer}; our approach can be applied to other
       frameworks as well.  We conduct experiments on training and inference using four
       CNNs (AlexNet,
 GoogLeNet, ResNet-50, and Inception-ResNet) and one RNN (seq2seq~\cite{seq2seq}).
 We find that our method not only reduces the memory consumption by up to 49.5\%
 but also accelerates propagation by up to a factor of four thanks to the rapidity
 of the optimized memory allocation and the ability to use larger mini-batch sizes.
\end{itemize}

One may raise a concern that a sample run for a profile can be
memory-inefficient and needs more memory than the physical capacity.  We can
obtain the profile even in such a case by utilizing an out-of-core
technique~\cite{vdnn,mem_tf} or Unified Memory in NVIDIA CUDA,
which enables us to run the model requiring memory over the capacity with
additional performance overhead, and then perform the succeeding runs without the overhead by disabling those techniques.

The rest of this paper is organized as follows.  We describe related work in
\sect{relwork} and introduce our approach in \sect{approach}.
\sect{impl} gives an implementation of our idea and explains how to
apply it to various DNNs.  \sect{exp} shows experimental results, and
\sect{concl} concludes this paper.

%% file: sec/relwork.tex
\section{Related work}
\label{sec:relwork}

The need to manage memory has emerged as DNNs have consumed huge memory. Many
deep learning frameworks---e.g., Theano~\cite{theano},
TensorFlow~\cite{tensorflow}, and Chainer~\cite{chainer}---allocate GPU memory
dynamically, based on memory pools, which are sets of unused memory blocks, and
garbage collection. Given a GPU
memory request, they find a memory block of an adequate size from a memory pool,
or allocate it from the physical memory if there is no such a block in the pool, and wrap
the block with a reference count.  When the memory block is reclaimed by the
garbage collection, it returns to the pool.  Our approach not only requires less
memory but also makes propagation faster thanks to no needs to search for
memory blocks dynamically, as discussed in \sect{exp}.

DNNs usually need much memory to store outputs from hidden layers for
backpropagation; we call such outputs \emph{intermediate results}.
MXNet~\cite{mxnet} analyzes a computational graph of a network model for memory
optimization.  While it optimizes memory usage of only intermediate results, our
approach can optimize the usage of, for example, temporary memory used to speed up
convolution, as well as memory for intermediate results.  Another advantage of
our approach is compatibility with dynamic memory allocation, which is discussed
in \sect{impl-gen}.
Chen et al.~\shortcite{recompute} and Meng et al.~\shortcite{mem_tf} reduce the
memory consumption for intermediate results by recomputing them as needed by
backpropagation.  Although the recomputation can reduce the memory consumption to
sublinear with respect to the number of layers, it needs an additional forward
propagation in every backpropagation.  Our approach never incurs such
performance overhead once the memory allocation is optimized.
Shirahata et al.~\shortcite{shirahata} reuse as many memory blocks allocated in
forward propagation as possible for backpropagation and develop an in-place
parameter update.  Their memory reduction method works only for training,
whereas our approach works effectively for both training and inference.

Another research direction for memory reduction is compression of DNNs.  The
work in this direction prunes connections of a network
model~\cite{prune,deepcompress} and/or quantizes learnable
parameters~\cite{quantize} in a way that preserves accuracy.
The compression approach seems complementary to our method, but a major difference
is that the compression changes the behavior of the model and so the compressed
model may not work as intended by network designers. On the other hand, our approach does
not change the computation involved by the model.  Another issue of the
compression is the need of time-consuming retraining~\cite{prune}.

Out-of-core algorithms, which offload device memory not used immediately to
a slower storage and prefetch it as necessary, are another way to run large
models on a device with limited memory.
Rhu et al.~\shortcite{vdnn} and Meng et al.~\shortcite{mem_tf} offload intermediate results on GPUs to CPU memory.
Although their work can run a very large model as long as the peak memory
consumption does not exceed the GPU memory capacity, it causes performance
degradation due to CPU-GPU communication for data transfer.
Unified Memory in NVIDIA CUDA allows more fine-grained offloading, but it incurs
significant and difficult-to-control overhead~\cite{mem_tf}.

Wang et al.~\shortcite{superneuron} integrate computational graph analysis,
out-of-core technology, and recomputation into one system, which has pros and cons of those methods inherently.
They focus on CNNs as a target, but it is not clear
how well their system works on other network models, such as RNNs.

Memory allocation generally can be regarded as \emph{a two dimensional strip packing problem (2SP)}.
It asks for a set of rectangular items to be placed in a container with a fixed width
and for the variable height to be minimized.
A memory block corresponds to a rectangular item with its allocation time as width and its memory size as height.
In this paper, we deal with a special case where the allocation times of all memory blocks
are fixed as input.
The problem is also known as the \emph{Dynamic Storage Allocation problem (DSA)}, a typical NP-hard problem~\cite{garey_johnson}.
2SP has been extensively studied theoretically and practically.
Steinberg~\shortcite{Steinberg1997} proposed a heuristic algorithm, whose approximation ratio is 2.
Arahori et al.~\shortcite{arahori2012} proposed a branch-and-bound-based exact algorithm to 2SP,
which works well for small and medium-sized instances.
Burk et al.~\shortcite{best-fit} proposed the \emph{best-fit algorithm},
which is a simple, constructive type of heuristics.
They showed that it works well for large-sized instances even compared with metaheuristics-based algorithms.

%% file: sec/approach.tex
\section{Profile-guided memory allocation}
\label{sec:approach}


From a profile of memory usage during hot propagation, we gather the information of
the memory blocks requested. Such information allows us to better determine
where to allocate the memory blocks in the physical memory.
We formulate the memory allocation problem as a special case of the two-dimensional
rectangle packing problems that is known as the Dynamic Storage Allocation (DSA).
We show the mixed integer programming (MIP) formulation of DSA, that
can be solved optimally for small instances, and then introduce a heuristic algorithm for
solving larger instances.


\subsection{Formulation}
\label{sec:formula}

We first introduce parameters of DSA.  We suppose that the number of
requested memory blocks, times when memory blocks are requested and released,
and sizes of memory blocks are given by the profile.  We also take the
available maximum memory size as a parameter.  Formally:

\begin{itemize}
 \item $n \in \mathbb{Z}$: number of memory blocks.
 \item $B = \{1,\dots,n\}$: a set of IDs of memory blocks.
 \item $W \in \mathbb{N}$: the available maximum memory size.
 \item $w_i \in \mathbb{N}$ $(i \in B)$: size of memory block $i$.
 \item $\underline{y_i} \in \mathbb{N}$ $(i \in B)$: time when $i$ is requested.
 \item $\overline{y_i} \in \mathbb{N}$ $(i \in B)$: time when $i$ is released.
\end{itemize}
We assume that these parameters do not change during the entire run (training
and inference) of a neural network.  This assumption is satisfied if the
propagation involved by the neural network is hot.  Many commonly used models satisfy this
condition. We give workarounds for network models where
only a part of the propagation is hot in \sect{impl-gen}.
A memory block $i$ is allocated during a time period $[\underline{y_j}, \overline{y_j})$;
we call the time period \emph{lifetime} of memory block $i$.

We next introduce the following decision variables of DSA.
\begin{itemize}
	\item $u \in \mathbb{Z}$: the peak memory usage.
	\item $x_i \in \mathbb{Z}$ $(i \in B)$: memory offset (or, starting
        address) of memory block $i$ within the entire allocated memory.
	\item $z_{ij} \in \{0, 1\}$ $((i, j) \in E)$: $0$ means that memory block $i$
        is located lower than block $j$ (i.e., $x_i + w_i \le x_j$) and $1$ means that
        it is not (i.e., $x_j + w_j \le x_i$).
\end{itemize}
We call the interval of memory address $[x_i, x_i + w_i)$ of memory block $i$ \emph{address space} of memory block $i$.

We assign memory offsets to memory blocks so that no two memory
blocks occupy the same address space at any given time. We do not need to check
for all pairs of memory blocks: it suffices to check those with overlapping lifetimes.
To this end, we introduce a notion of \emph{possible colliding pairs}:
\[
 E = \{(i, j) \in B \times B \mid i < j \; \text{and} \;
   [\underline{y_i}, \overline{y_i}) \cap [\underline{y_j}, \overline{y_j}) \ne \emptyset\},
\]
which is a set of memory block pairs that have overlapping lifetimes.
Note that any two memory blocks not in $E$ do not share the same address space at the same time because their lifetimes do not overlap.

The objective of DSA is to minimize the peak memory usage.
We formulate DSA in the form of a MIP as follows:
\begin{align}
\text{min} \quad& u,\\
\text{s.t.} \quad
& x_i + w_i \leq u, \; i \in B, \\
& x_i + w_i \leq x_j + z_{ij} W, \; (i, j) \in E, \\
& x_j + w_j \leq x_i + (1 - z_{ij}) W, \; (i, j) \in E, \\
& 0 \leq u \leq W, \; \text{(5)} \qquad  \setcounter{equation}{5}
  x_i \geq 0, \; i \in B.
\end{align}
Equations (1) and (2) represent the minimization of the peak memory usage.
Equations (3) and (4) denote the non-overlapping constraints of possible colliding pairs.
We can exactly solve the above MIP with CPLEX for small instances.

\begin{figure}[t]
 \centering
 \begin{subfigure}[t]{.29\linewidth}
  \includegraphics[width=\linewidth]{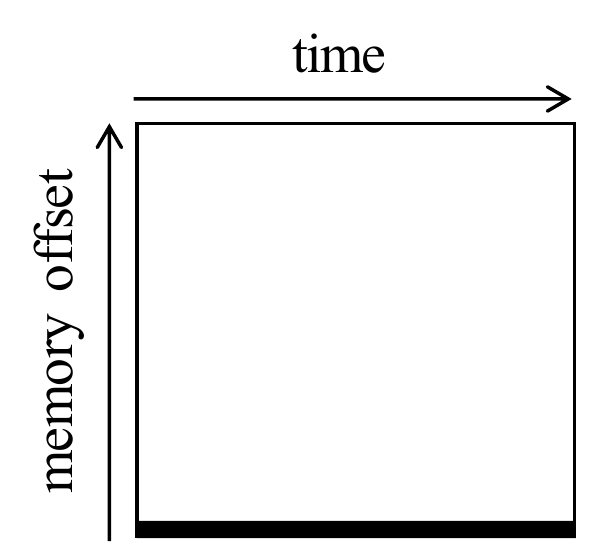}
  \caption{The initial state.}
  \label{fig:rect-a}
 \end{subfigure}
 \hfil
 \begin{subfigure}[t]{.22\linewidth}
  \includegraphics[width=\linewidth]{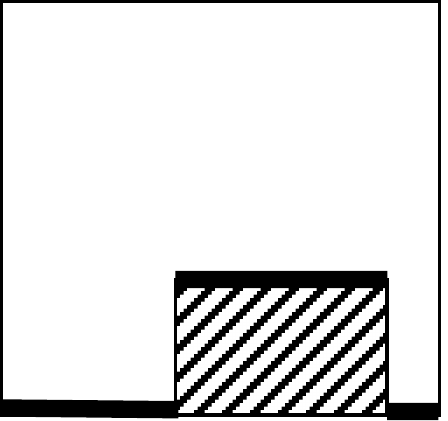}
  \caption{Placing the first block.}
  \label{fig:rect-b}
 \end{subfigure}
 \hfil
 \begin{subfigure}[t]{.22\linewidth}
  \includegraphics[width=\linewidth]{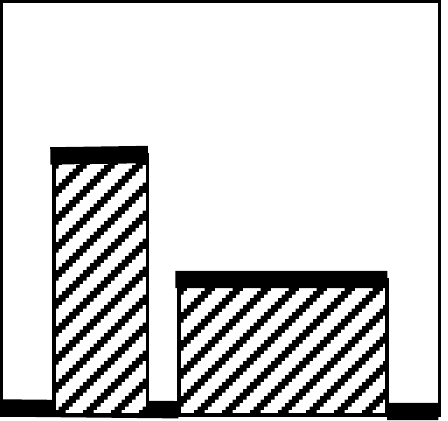}
  \caption{Placing the second block.}
  \label{fig:rect-c}
 \end{subfigure}
 \hfil
 \begin{subfigure}[t]{.22\linewidth}
  \includegraphics[width=\linewidth]{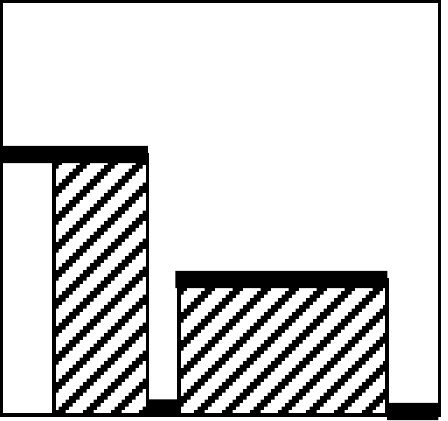}
  \caption{Lifting up the lowest offset.}
  \label{fig:rect-d}
 \end{subfigure}
 \caption{A running example of the best-fit heuristic.}
 \label{fig:rect}
\end{figure}

\subsection{Best-fit heuristic}
\label{sec:heuristics}
We design our heuristic to DSA on the basis of the best-fit heuristic~\cite{best-fit} to 2SP.
In \fig{rect}, the $x$-axis and $y$-axis denote times and memory
offsets, respectively. Note that the objective is to place all memory blocks in the
rectangle so that the top block is placed as low as possible.
The heuristic repeats two operations
until all memory blocks are placed: (1) choosing an offset and (2)
searching for a memory block that can be placed at the chosen offset without
colliding with memory blocks placed already.
We illustrate how our heuristic works via an example shown in \fig{rect}.  In
the beginning of the heuristic (\fig{rect-a}), since no memory blocks are
placed, we choose zero as the memory offset.  When searching for a memory block,
we always choose a block with the longest lifetime among blocks that can be
placed at the considered offset.  In the case of \fig{rect}, we choose the
memory block that has the longest lifetime and place it (\fig{rect-b}).  After the
placement, there are three candidates of a memory offset (the bold lines in
\fig{rect}; we call them offset lines).  When there are multiple offsets, we
always choose the lowest one (if there are multiple lowest offsets, the leftmost
one is chosen).  Next, we search for a memory block for the chosen offset and
place it (\fig{rect-c}).  If there are no blocks that can be placed at the
chosen offset, we ``lift up'' the line for the offset by merging it with the
lowest adjacent offset line as in \fig{rect-d} (if offsets of adjacent lines are the same,
it is merged with both) and again choose an offset and search for a memory
block.
\iffull
\else
The computational time complexity of our heuristic is quadratic in the number of memory blocks.
\fi

%% file: sec/impl.tex
\section{Implementation}
\label{sec:impl}

We incorporate the best-fit heuristic in Chainer to optimize the GPU memory
usage.  This section describes the details including how to apply our
approach to \emph{any} network models.

\subsection{Memory profiling}
\label{sec:impl-monitor}

Since Chainer allocates memory blocks at run time, we profile GPU memory usage
by monitoring memory allocation and free operations in a sample run.  To obtain
memory request time $\underline{y_i}$ and release time $\overline{y_i}$, we use
a global integer variable $y$, which represents the current time and is
increased after each allocation and free.  We also have a global integer
variable $\id$ that denotes the ID of the next requested memory block.

Given a sample input, we initialize the global variables with one and run the
model with the input.  When receiving a request with memory size $s$, we extend
$B$ (the set of memory block IDs) with \id, set $s$ and $y$ to $w_\id$ and
$\underline{y_\id}$, respectively, and finally increase $\id$ and $y$.  When
memory block $i$ is released, $y$ is set to $\overline{y_i}$ and then increased.

\subsection{Memory allocation}
\label{sec:impl-mem-alloc}

After obtaining the parameters from the sample run, we calculate the peak memory usage $u$
and memory offsets $x_i$ for memory blocks $i$ by solving DSA and
then allocate GPU memory of size $u$; we write $p$ for the address of the
memory.  In the rest of the running of the model, we return memory address $p + x_i$
for a request of memory block $i$.  We identify memory blocks by maintaining the global
variable \id, which is initialized with one before starting each forward
propagation.  When a memory block is requested, we return address $p + x_\id$
and increase \id.  This is sound since the propagation should be computed in
the same way as in the sample run, where $\id$ is always increased after each
allocation.

\subsection{Generalization for non-hot propagation}
\label{sec:impl-gen}

The memory allocation in \sect{impl-mem-alloc} is unsound for models which, for
different inputs, (1) perform non-hot propagation (that is, it is computed
differently) and (2) request memory of different sizes.  This section gives
workarounds to avoid them.

A workaround for the first issue is very simple: we do \emph{not} optimize the
usage of memory requested in the non-hot part of the propagation.
To this end, we provide two operations, \texttt{interrupt} and \texttt{resume},
which interrupt and resume the monitoring of memory operations, respectively.
When entering a non-hot part,
we call \texttt{interrupt}; and, when leaving that part, we call \texttt{resume}.
Since our method optimizes only the profiled part of memory usage, the memory
requested between calls to \texttt{interrupt} and \texttt{resume} is out of
the scope of the optimization.

The second issue is resolved by \emph{reoptimization}.  In this approach, we
continue the monitoring of memory operations after optimizing the memory usage
and, when detecting a request for larger memory than expected, we
reoptimize the memory allocation by using the new observed parameters---note that
we do not need reoptimization for requests of smaller memory.  This workaround may
incur an additional performance cost, but it is very low as shown in
\sect{exp-seq2seq}.

%% file: sec/exp.tex
\section{Experiments}
\label{sec:exp}

\begin{figure}[t!]
 \centering
 \begin{subfigure}{\linewidth}
   \includegraphics[width=\linewidth]{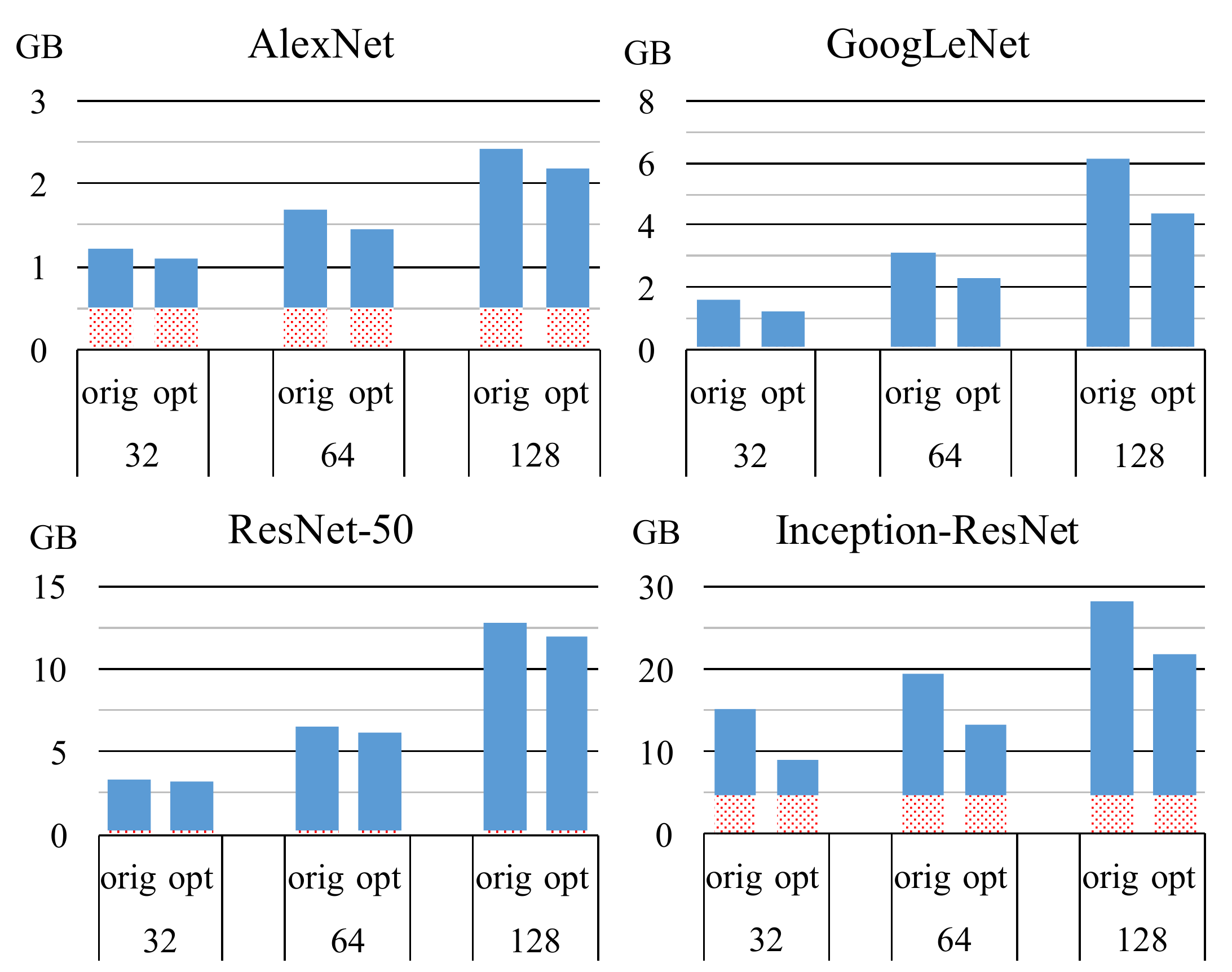}
  \caption{CNN training.}
  \label{fig:exp-cnn-mem-train}
 \end{subfigure}
 \begin{subfigure}{\linewidth}
   \includegraphics[width=\linewidth]{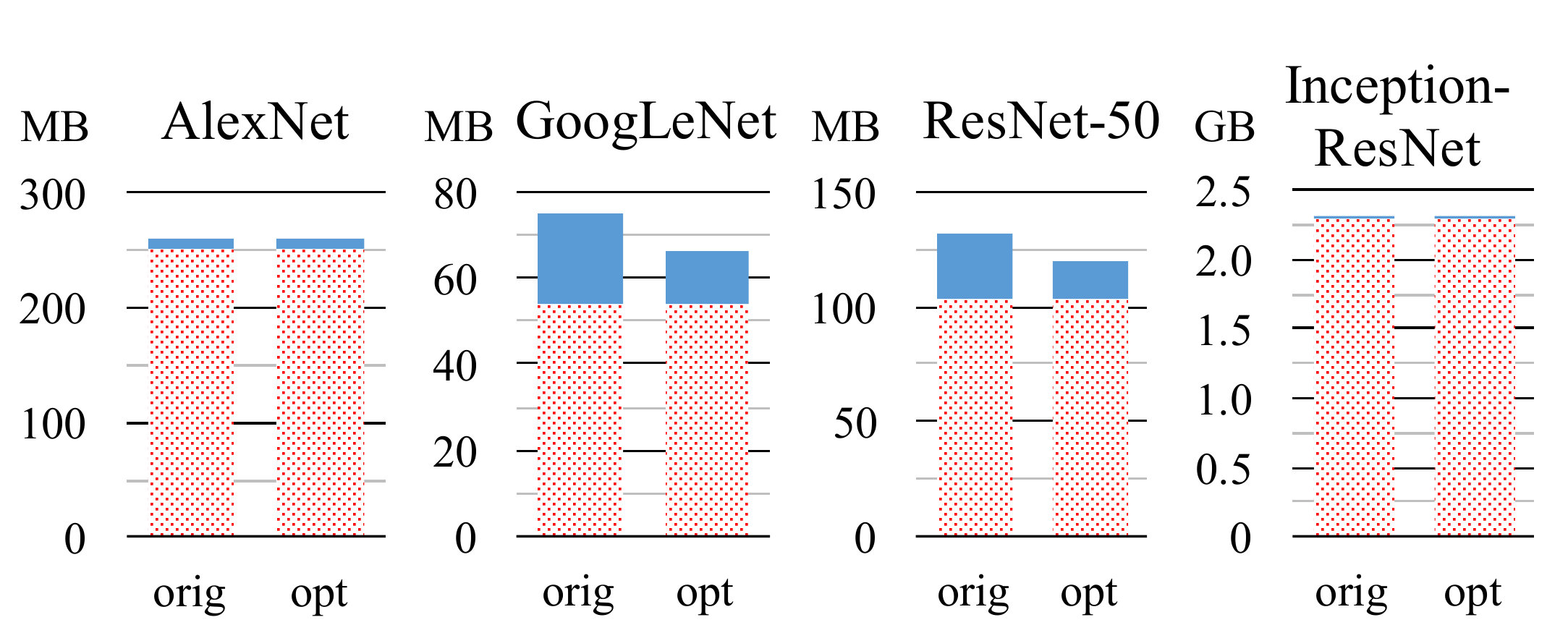}
  \caption{CNN inference.}
  \label{fig:exp-cnn-mem-inf}
 \end{subfigure}
 \begin{subfigure}[b]{.68\linewidth}
   \includegraphics[width=\linewidth]{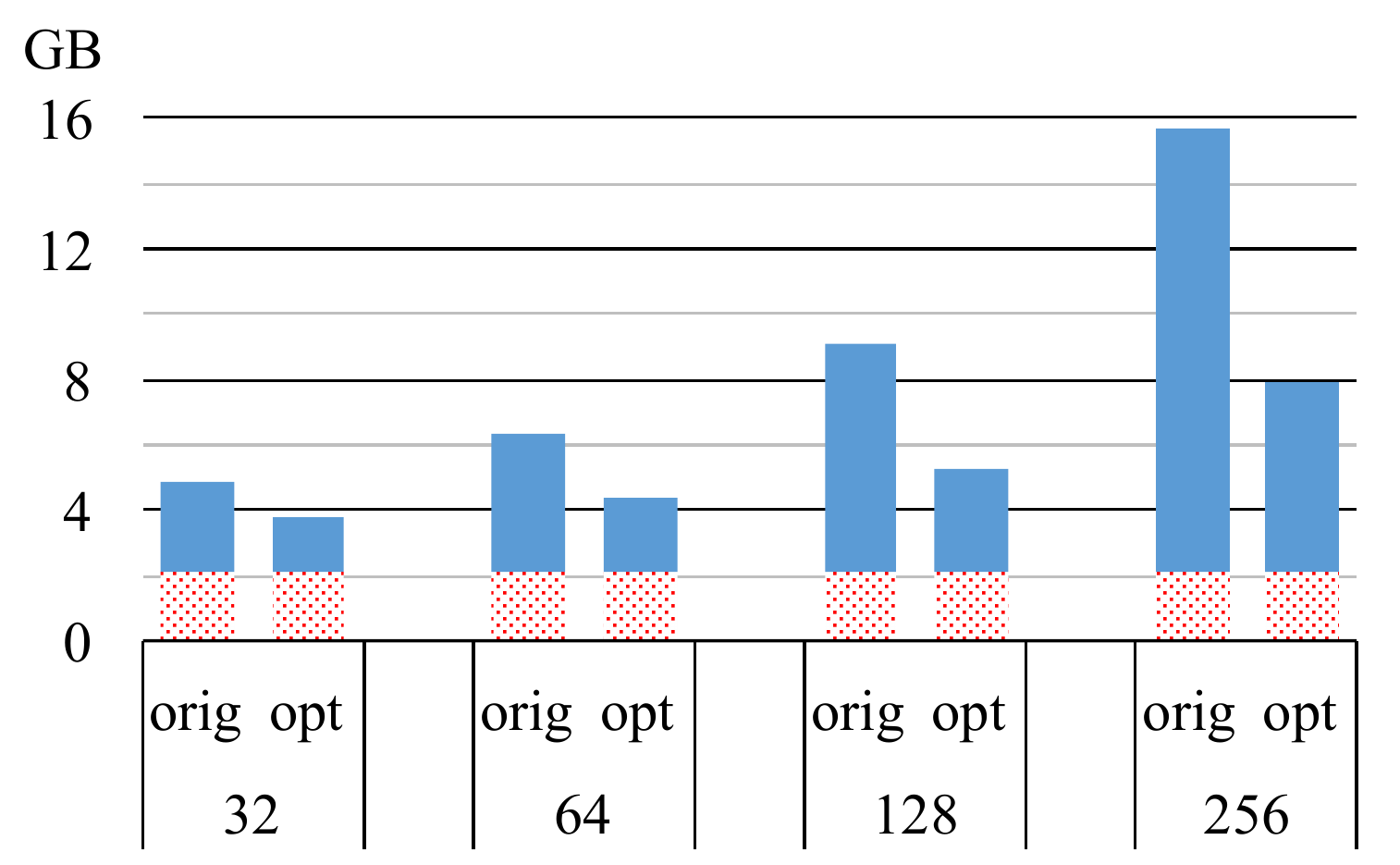}
  \caption{Seq2Seq training.}
  \label{fig:exp-rnn-mem-train}
 \end{subfigure}
 \hspace{.3ex}
 \begin{subfigure}[b]{.29\linewidth}
  \includegraphics[width=\linewidth]{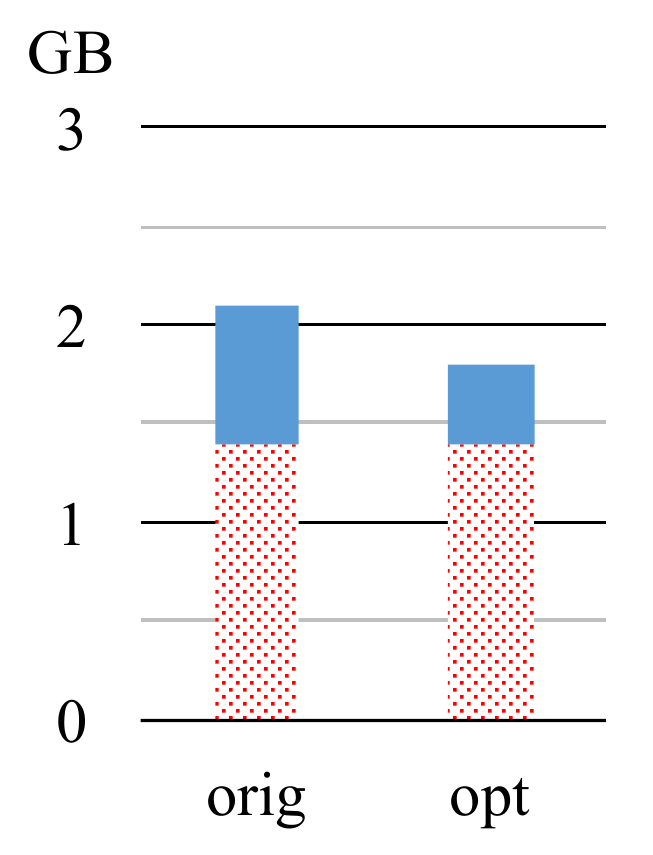}
  \caption{Seq2Seq inference.}
  \label{fig:exp-rnn-mem-inf}
 \end{subfigure} \\
 \vspace{2ex}
 \begin{subfigure}{.85\linewidth}
  \includegraphics[width=\linewidth]{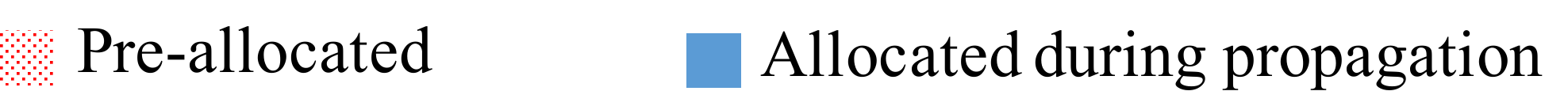}
 \end{subfigure}
 \caption{The memory consumption.}
 \label{fig:exp-mem}
\end{figure}

\begin{figure}[t!]
  \centering
 \begin{subfigure}{\linewidth}
   \includegraphics[width=\linewidth]{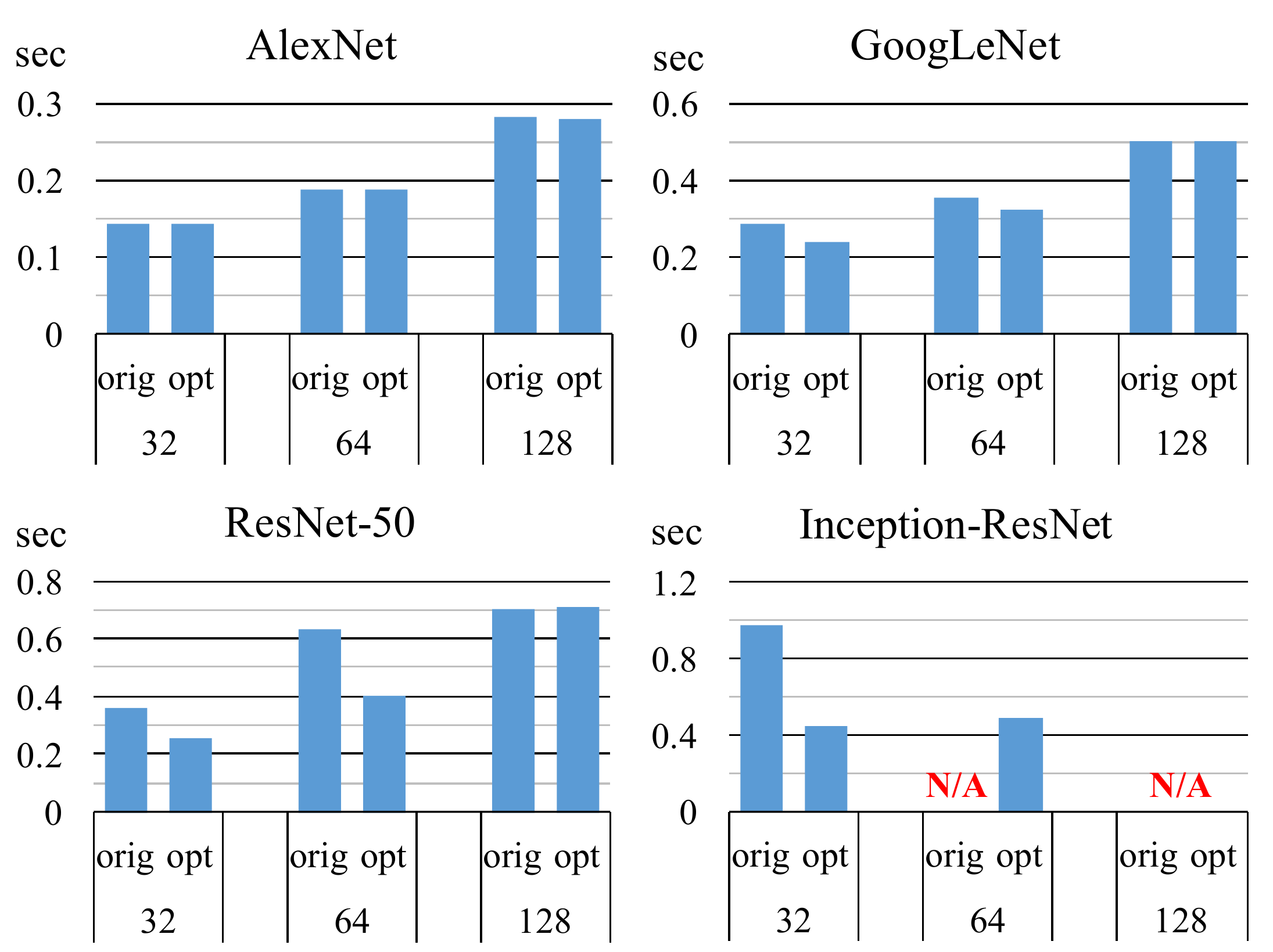}
  \caption{CNN training.}
  \label{fig:exp-cnn-run-train}
 \end{subfigure}
 \begin{subfigure}{\linewidth}
   \includegraphics[width=\linewidth]{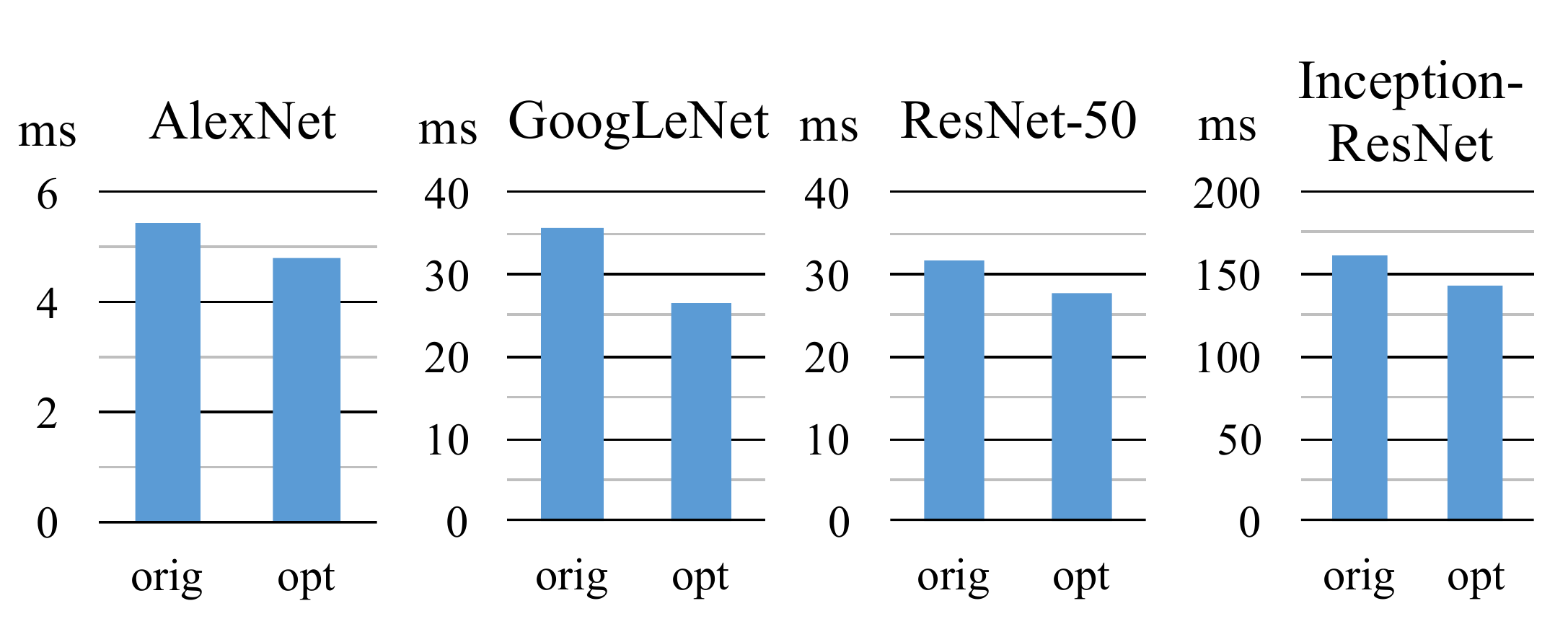}
  \caption{CNN inference.}
  \label{fig:exp-cnn-run-inf}
 \end{subfigure}
 \begin{subfigure}{.72\linewidth}
   \includegraphics[width=\linewidth]{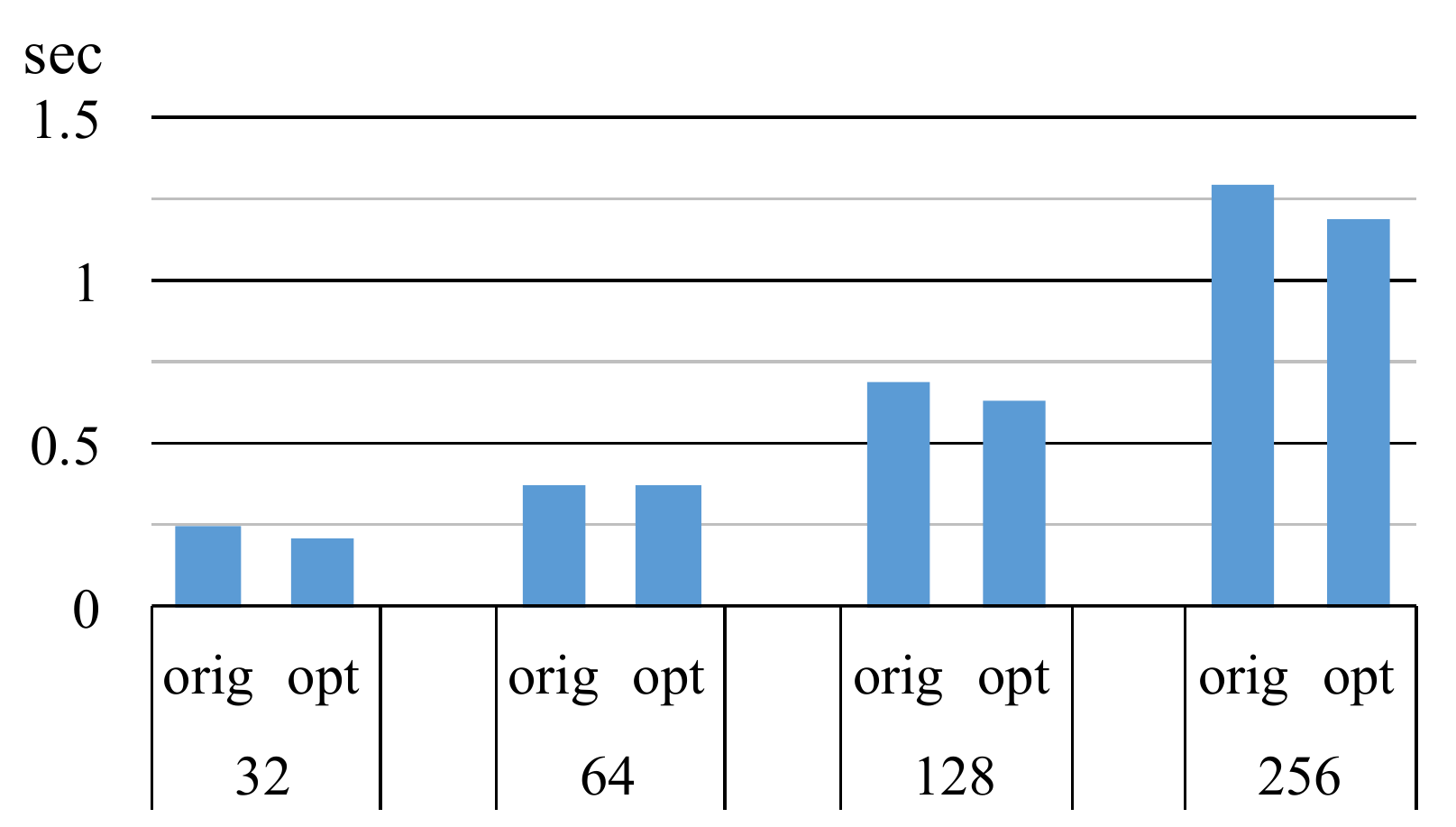}
  \caption{Seq2Seq training.}
  \label{fig:exp-rnn-run-train}
 \end{subfigure}
 \hspace{.3ex}
 \begin{subfigure}{.24\linewidth}
   \includegraphics[width=\linewidth]{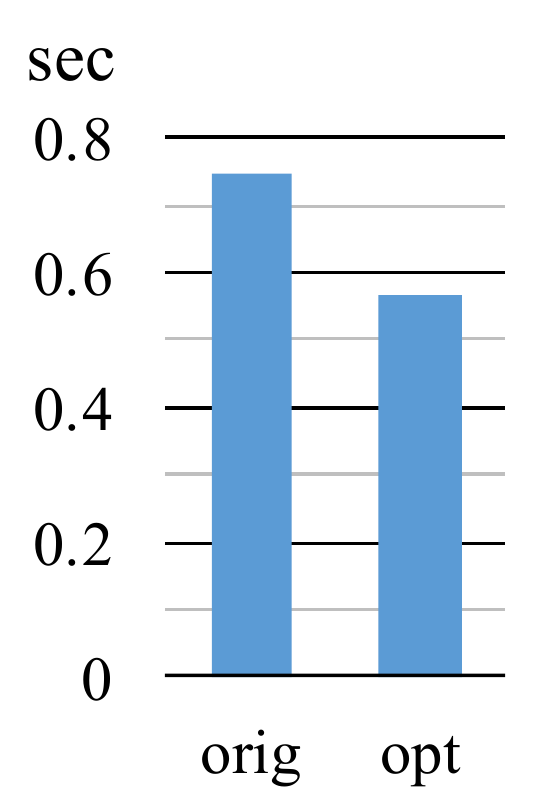}
  \caption{Seq2Seq inference.}
  \label{fig:exp-rnn-run-inf}
 \end{subfigure}
 \caption{The average elapsed times of processing one mini-batch in training
 and one input data in inference.}
 \label{fig:exp-run}
\end{figure}


\begin{figure}[t]
 \centering
 \begin{subfigure}{\linewidth}
  \includegraphics[width=\linewidth]{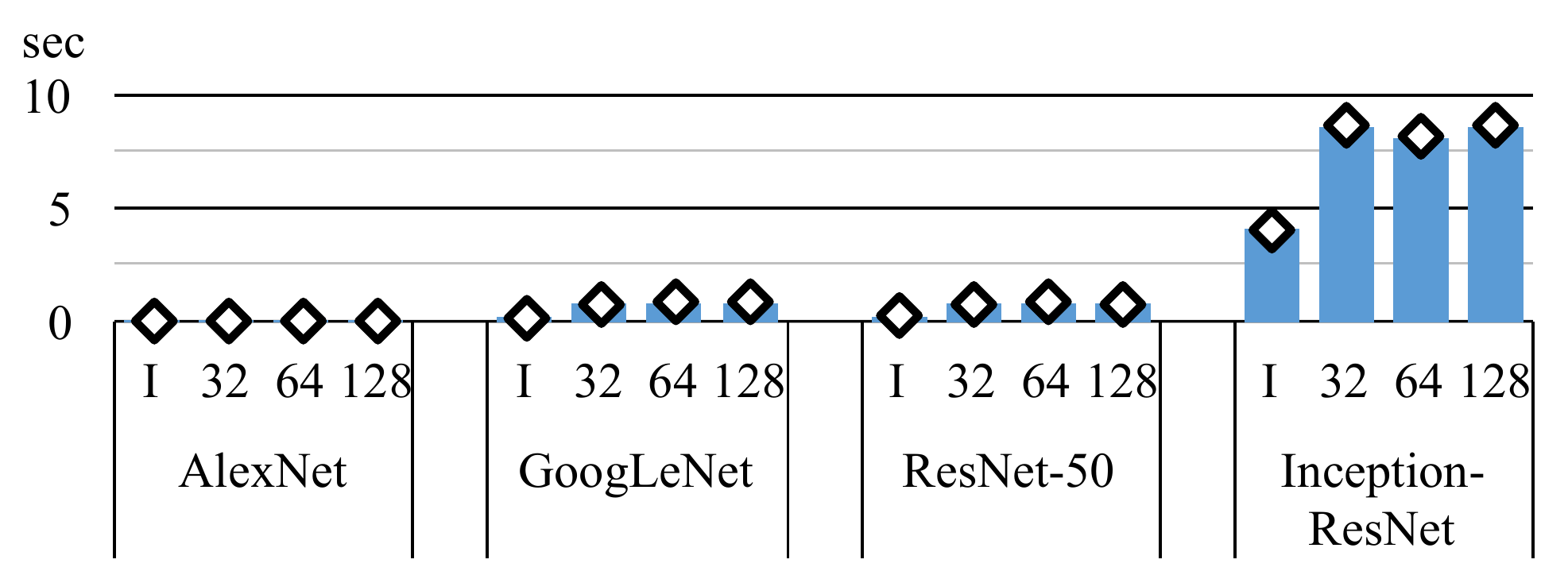}
  \caption{CNNs.}
  \label{fig:exp-cnn-heuristics}
 \end{subfigure}
 \begin{subfigure}{\linewidth}
  \includegraphics[width=\linewidth]{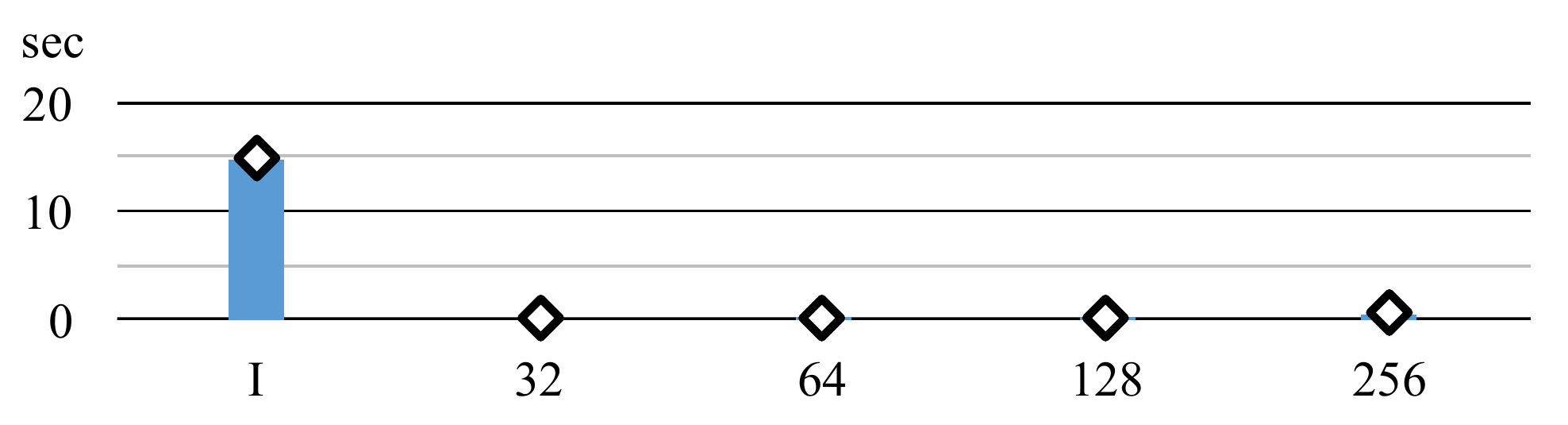}
  \caption{Seq2Seq.}
  \label{fig:exp-rnn-heuristics}
 \end{subfigure}
 \caption{The running times of the best-fit heuristic. ``I'' on the x-axes
 means that the corresponding numbers are the times for the inference and
 32, 64, 128, and 256 denote mini-batch sizes in the training.}
 \label{fig:exp-heuristics}
\end{figure}

\subsection{Configurations}

We compare the GPU device memory consumption (\fig{exp-mem}) and the running
times to process one mini-batch in training and inference (\fig{exp-run}) in
Chainer (version 3 RC 1.0), which is a baseline and denoted by \emph{orig} in
figures for shorthand, and our optimized version, denoted by \emph{opt}, on four
CNNs (AlexNet, GoogLeNet, ResNet-50, and Inception-ResNet), and one RNN (seq2seq).
%
Training of the CNNs is performed with 32, 64, and 128 mini-batch sizes, and that of seq2seq is with
32, 64, 128, and 256 ones.  Inference performs only forward propagation for one
input data.  We use ImageNet~\cite{ILSVRC15} and the English-French corpus from
WMT15\footnote{\url{http://www.statmt.org/wmt15/}} as datasets for the CNNs and
seq2seq, respectively.  We use the first 1000 training mini-batches for the
warm-up and next 2000 mini-batches for the evaluation.
We turn on Unified Memory of NVIDIA CUDA, which allows us to run models
requiring more memory than the physical capacity, in the experiments for
memory consumption but turn it off in the measurement of running times since it
may incur performance overhead.

We also evaluate the best-fit heuristic implemented in Python in two experiments.
We first compare the solutions by the heuristic with the optimal solutions found by CPLEX version 12.8 within one hour.
We also compare the computation time of the heuristic for different configurations (\fig{exp-heuristics}).

All experiments are run on an IBM POWER8 machine with two 4GHz 10-core POWER8
processors, 512 GB RAM, and NVIDIA Tesla P100 GPUs equipped with 16 GB device
memory.  Options except mini-batch sizes follow the scripts provided by
Chainer.

Finally, we make a few remarks.
The first is on the GPU memory management system of our baseline.  The original
Chainer uses memory pools for memory reuse, as described in \sect{relwork},
and reduces the memory consumption somewhat compared with naive, network-wise
memory allocation, which always allocates a memory block from the physical
device memory for each request.  For example, we observed that, in the training
of AlexNet with 32 mini-batch size, the network-wise memory allocation consumes
1.50 GB device memory whereas the pool-based memory allocation does only 1.21 GB
memory.  In this section, we show that our approach achieves reduction of more
memory than the pool-based method.
The second remark is on convolution algorithms.  There are many algorithms for
computing convolution.  The most memory-efficient algorithm needs
memory only for inputs and outputs, but we can calculate the convolution much
faster by allocating additional temporary memory, called \emph{workspace}.  Although
the optimized version could be accelerated by allocating larger workspace than
the original Chainer, the experiments use workspace of the same size (8 MB by
default) in both versions for comparing only the effect of the memory
optimization.

\subsection{CNNs}

\subsubsection{Training}

The total memory consumption during the training of CNNs is shown in
\fig{exp-cnn-mem-train}, where the amount of memory retained in the entire
training (e.g., memory for learnable parameters and gradients) is indicated by
doted red bars and the amount of memory released until the end of each
propagation is indicated by solid blue bars; our method optimizes usage of only the
latter.  \fig{exp-cnn-mem-train} shows that our optimization works well in all
models and is the most effective for Inception-ResNet.  Specifically, in 64
mini-batch size, the memory consumption in the optimized version fits within the
physical memory capacity (16 GB), whereas the required memory in the original
Chainer exceeds the capacity considerably.

The average training times per mini-batch are reported in \fig{exp-cnn-run-train}, where
``N/A'' means that we could not train the model due to insufficient memory.
The results indicate that our approach accelerates
the training of some models (GoogLeNet, ResNet-50, and Inception-ResNet) even
using the same mini-batch size.  This
is because the optimized version allocates memory quite quickly.  The original
Chainer, given a memory request, searches for an available, sufficiently sized
memory block from a memory pool, and the running cost of this memory search
increase as the number of memory blocks in the pool increases.
In contrast, the optimized version just returns a memory address calculated
before the training.  This is significant especially in Inception-ResNet, for
which the training in the optimized version is 2.19 times as fast as in the
original.  This advantage however may be hidden behind GPU computation in a
large mini-batch size, as in GoogLeNet and ResNet-50 with 128 mini-batch size.
\fig{exp-cnn-run-train} also shows that our method allows use of a larger
mini-batch size, which may enable us to utilize GPUs more fully.  For example,
the training of Inception-ResNet in the optimized version (64 mini-batch size)
utilizes more GPU cores than that in the original (32 mini-batch size),
and the number of images processed per second by the former is 3.95 times as
large as that by the latter.

\subsubsection{Inference}

The memory consumption in inference is shown in \fig{exp-cnn-mem-inf}.  Since
the inference does not need to retain memory for intermediate results, most
memory blocks can be reused even in the pool-based memory management of
Chainer.  Nevertheless, we successfully reduce the total memory amounts in
GoogLeNet and ResNet-50 by 12.6\% and 10.0\%, respectively.  As for running time
performance, inference is accelerated in all models, as shown in
\fig{exp-cnn-run-inf}, because the GPU computation for inference is lightweight
and the cost of search for memory blocks is more dominant.

\subsubsection{Heuristic}

%
CPLEX could obtain the optimal solutions only in two configurations (inference
using AlexNet and GoogLeNet), and the objective function values by the heuristic
and CPLEX match (10169344 and 12202496, respectively).  Our heuristic thus works
very well at least in small instances.
The execution times of the heuristic are shown in \fig{exp-cnn-heuristics},
which indicates that the heuristic works quickly enough for practical use.

\subsection{Seq2Seq}
\label{sec:exp-seq2seq}

\subsubsection{Training and inference}

\fig{exp-rnn-mem-train} shows the memory consumption immediately after
processing 10 mini-batches in the training of seq2seq and demonstrates that our
approach significantly reduces the memory consumption.  In the original Chainer,
since the training of seq2seq requires differently sized memory for different
inputs, memory blocks allocated in a training loop may not be used in the
succeeding loops, and the whole of such unused blocks finally reaches the device
memory capacity.  In contrast, we recompute how to allocate
memory when necessary, which allows us to keep the memory consumption as
low as possible.

The growth of unused memory blocks causes degradation of the running time
performance.  \fig{exp-rnn-run-train}, which gives the running times of training
seq2seq, shows that the optimized version is faster in 32 mini-batch size and
the original Chainer catches up in 64 mini-batch size for the same reason as in
CNNs.  In 128 and 256 mini-batch sizes, however, the training in the optimized
version becomes faster than that in the original.  We guess that this is due to the
waste of GPU memory by the pool-based memory management system.
When requested memory cannot be allocated due to insufficient
free memory, the pool-based memory management system frees all unused memory
blocks and allocates subsequent requested memory from the physical GPU memory,
which has higher run-time cost than reusing memory blocks in a pool.  Our memory allocation
does not use memory pools except a small part enclosed by \texttt{interrupt} and
\texttt{resume}, so we rarely need to free memory.  While our approach instead must recompute
memory addresses when the requested memory is larger than expected
(\sect{impl-gen}), the recomputation cost in the training is low, as shown in
\fig{exp-rnn-heuristics}, and the recomputation becomes less frequent as the
training proceeds.

As for the inference, the amount of consumed memory and the running time reduce by 14.6\%
(\fig{exp-rnn-mem-inf}) and 23.8\% (\fig{exp-rnn-run-inf}), respectively.

\subsubsection{Heuristic}

As shown in \fig{exp-rnn-heuristics}, the heuristic algorithm takes much
longer in the inference, whereas it terminates quickly for the training
formulas.  This is due to the Chainer script that we use for the evaluation: the
script always generates 100 words for inference, whereas it cuts sentences used
for the training into up to 50 words.  Thus, the inference requests many more
memory blocks than the training, and the heuristic takes long in the inference.
Fortunately, this should not be problematic in
practice, because we can solve DSA with idle CPUs after responding to an
inference request.
We note that the running time performance of the heuristic can be improved by
using faster languages, such as C and {\cpp}.
CPLEX could not obtain the optimal solutions within the 1-hour time limit.

%% file: sec/conclusion.tex
\section{Conclusion}
\label{sec:concl}

We propose a profile-guided memory optimization for DNNs.
We develop a simple heuristic algorithm to DSA to obtain efficient and fast memory allocation,
and incorporate the heuristic in Chainer.  We experimentally confirmed that our
method reduces the memory consumption and accelerates
propagation in both training and inference using CNNs and seq2seq.

%% file: main.bbl
\begin{thebibliography}{}

\bibitem[\protect\citeauthoryear{Abadi \bgroup \em et al.\egroup
  }{2016}]{tensorflow}
Mart{\'{\i}}n Abadi, Paul Barham, Jianmin Chen, Zhifeng Chen, Andy Davis,
  Jeffrey Dean, Matthieu Devin, Sanjay Ghemawat, Geoffrey Irving, Michael
  Isard, Manjunath Kudlur, Josh Levenberg, Rajat Monga, Sherry Moore, Derek~G.
  Murray, Benoit Steiner, Paul~A. Tucker, Vijay Vasudevan, Pete Warden, Martin
  Wicke, Yuan Yu, and Xiaoqiang Zheng.
\newblock {TensorFlow}: A system for large-scale machine learning.
\newblock In {\em Proc.\ of OSDI}, pages 265--283, 2016.

\bibitem[\protect\citeauthoryear{Arahori \bgroup \em et al.\egroup
  }{2012}]{arahori2012}
Yohei Arahori, Takashi Imamichi, and Hiroshi Nagamochi.
\newblock An exact strip packing algorithm based on canonical forms.
\newblock {\em Computers \& Operations Research}, 39(12):2991--3011, 2012.

\bibitem[\protect\citeauthoryear{Bastien \bgroup \em et al.\egroup
  }{2012}]{theano}
Fr{\'{e}}d{\'{e}}ric Bastien, Pascal Lamblin, Razvan Pascanu, James Bergstra,
  Ian~J. Goodfellow, Arnaud Bergeron, Nicolas Bouchard, David Warde{-}Farley,
  and Yoshua Bengio.
\newblock {Theano}: new features and speed improvements.
\newblock {\em CoRR}, abs/1211.5590, 2012.

\bibitem[\protect\citeauthoryear{Burke \bgroup \em et al.\egroup
  }{2004}]{best-fit}
Edmund~K. Burke, Graham Kendall, and Glenn Whitwell.
\newblock A new placement heuristic for the orthogonal stock-cutting problem.
\newblock {\em Operations Research}, 52(4):655--671, 2004.

\bibitem[\protect\citeauthoryear{Chen \bgroup \em et al.\egroup }{2015}]{mxnet}
Tianqi Chen, Mu~Li, Yutian Li, Min Lin, Naiyan Wang, Minjie Wang, Tianjun Xiao,
  Bing Xu, Chiyuan Zhang, and Zheng Zhang.
\newblock {MXNet}: A flexible and efficient machine learning library for
  heterogeneous distributed systems.
\newblock {\em CoRR}, abs/1512.01274, 2015.

\bibitem[\protect\citeauthoryear{Chen \bgroup \em et al.\egroup
  }{2016}]{recompute}
Tianqi Chen, Bing Xu, Chiyuan Zhang, and Carlos Guestrin.
\newblock Training deep nets with sublinear memory cost.
\newblock {\em CoRR}, abs/1604.06174, 2016.

\bibitem[\protect\citeauthoryear{Garey and Johnson}{1979}]{garey_johnson}
M.~R. Garey and D.~S. Johnson.
\newblock {\em Computers and Intractability : A Guide to the Theory of
  {NP}-Completeness}.
\newblock Series of Books in the Mathematical Sciences. W. H. Freeman, 1979.

\bibitem[\protect\citeauthoryear{Gong \bgroup \em et al.\egroup
  }{2014}]{quantize}
Yunchao Gong, Liu Liu, Ming Yang, and Lubomir~D. Bourdev.
\newblock Compressing deep convolutional networks using vector quantization.
\newblock {\em CoRR}, abs/1412.6115, 2014.

\bibitem[\protect\citeauthoryear{Han \bgroup \em et al.\egroup }{2015}]{prune}
Song Han, Jeff Pool, John Tran, and William~J. Dally.
\newblock Learning both weights and connections for efficient neural network.
\newblock In {\em Proc.\ of NIPS}, pages 1135--1143, 2015.

\bibitem[\protect\citeauthoryear{Han \bgroup \em et al.\egroup
  }{2016}]{deepcompress}
Song Han, Huizi Mao, and William~J. Dally.
\newblock Deep compression: Compressing deep neural network with pruning,
  trained quantization and huffman coding.
\newblock In {\em Proc.\ of ICLR}, 2016.

\bibitem[\protect\citeauthoryear{He \bgroup \em et al.\egroup }{2016}]{resnet}
Kaiming He, Xiangyu Zhang, Shaoqing Ren, and Jian Sun.
\newblock Deep residual learning for image recognition.
\newblock In {\em Proc.\ of CVPR}, pages 770--778, 2016.

\bibitem[\protect\citeauthoryear{Hochreiter and Schmidhuber}{1997}]{lstm}
Sepp Hochreiter and J{\"{u}}rgen Schmidhuber.
\newblock Long short-term memory.
\newblock {\em Neural Computation}, 9(8):1735--1780, 1997.

\bibitem[\protect\citeauthoryear{Krizhevsky \bgroup \em et al.\egroup
  }{2012}]{alexnet}
Alex Krizhevsky, Ilya Sutskever, and Geoffrey~E. Hinton.
\newblock Imagenet classification with deep convolutional neural networks.
\newblock In {\em Proc.\ of NIPS}, pages 1106--1114, 2012.

\bibitem[\protect\citeauthoryear{LeCun \bgroup \em et al.\egroup
  }{2015}]{DL-Nature}
Yann LeCun, Yoshua Bengio, and Geoffrey~E. Hinton.
\newblock Deep learning.
\newblock {\em Nature}, 521(7553):436--444, 2015.

\bibitem[\protect\citeauthoryear{Meng \bgroup \em et al.\egroup
  }{2017}]{mem_tf}
Chen Meng, Minmin Sun, Jun Yang, Minghui Qiu, and Ynag Gu.
\newblock Training deeper models by {GPU} memory optimization on {TensorFlow}.
\newblock In {\em Proc.\ of ML Systems Workshop in NIPS}, 2017.

\bibitem[\protect\citeauthoryear{Rhu \bgroup \em et al.\egroup }{2016}]{vdnn}
Minsoo Rhu, Natalia Gimelshein, Jason Clemons, Arslan Zulfiqar, and Stephen~W.
  Keckler.
\newblock {vDNN}: Virtualized deep neural networks for scalable,
  memory-efficient neural network design.
\newblock In {\em Proc.\ of MICRO}, pages 1--13, 2016.

\bibitem[\protect\citeauthoryear{Russakovsky \bgroup \em et al.\egroup
  }{2015}]{ILSVRC15}
Olga Russakovsky, Jia Deng, Hao Su, Jonathan Krause, Sanjeev Satheesh, Sean Ma,
  Zhiheng Huang, Andrej Karpathy, Aditya Khosla, Michael Bernstein,
  Alexander~C. Berg, and Li~Fei-Fei.
\newblock {ImageNet Large Scale Visual Recognition Challenge}.
\newblock {\em International Journal of Computer Vision}, 115(3):211--252,
  2015.

\bibitem[\protect\citeauthoryear{Shirahata \bgroup \em et al.\egroup
  }{2016}]{shirahata}
Koichi Shirahata, Yasumoto Tomita, and Atsushi Ike.
\newblock Memory reduction method for deep neural network training.
\newblock In {\em Proc.\ of MLSP}, pages 1--6, 2016.

\bibitem[\protect\citeauthoryear{Steinberg}{1997}]{Steinberg1997}
A.~Steinberg.
\newblock A strip-packing algorithm with absolute performance bound 2.
\newblock {\em SIAM Journal on Computing}, 26:401--409, 1997.

\bibitem[\protect\citeauthoryear{Sutskever \bgroup \em et al.\egroup
  }{2014}]{seq2seq}
Ilya Sutskever, Oriol Vinyals, and Quoc~V. Le.
\newblock Sequence to sequence learning with neural networks.
\newblock In {\em Proc.\ of NIPS}, pages 3104--3112, 2014.

\bibitem[\protect\citeauthoryear{Szegedy \bgroup \em et al.\egroup
  }{2015}]{googlenet}
Christian Szegedy, Wei Liu, Yangqing Jia, Pierre Sermanet, Scott~E. Reed,
  Dragomir Anguelov, Dumitru Erhan, Vincent Vanhoucke, and Andrew Rabinovich.
\newblock Going deeper with convolutions.
\newblock In {\em Proc.\ of CVPR}, pages 1--9, 2015.

\bibitem[\protect\citeauthoryear{Szegedy \bgroup \em et al.\egroup
  }{2017}]{incept-resnet}
Christian Szegedy, Sergey Ioffe, Vincent Vanhoucke, and Alexander~A. Alemi.
\newblock {Inception}-v4, {Inception-ResNet} and the impact of residual
  connections on learning.
\newblock In {\em Proc.\ of AAAI}, pages 4278--4284, 2017.

\bibitem[\protect\citeauthoryear{Tokui \bgroup \em et al.\egroup
  }{2015}]{chainer}
Seiya Tokui, Kenta Oono, Shohei Hido, and Justin Clayton.
\newblock {Chainer}: A next-generation open source framework for deep learning.
\newblock In {\em Proc.\ of Workshop on Machine Learning Systems in NIPS},
  2015.

\bibitem[\protect\citeauthoryear{Wang \bgroup \em et al.\egroup
  }{2018}]{superneuron}
Linnan Wang, Jinmian Ye, Yiyang Zhao, Wei Wu, Ang Li, Shuaiwen~Leon Song,
  Zenglin Xu, and Tim Kraska.
\newblock {SuperNeurons}: Dynamic {GPU} memory management for training deep
  neural networks.
\newblock {\em CoRR}, abs/1801.04380, 2018.

\end{thebibliography}
